\documentclass[final]{svjour3}
\usepackage{graphicx}
\usepackage{rotating}
\usepackage{amssymb}
\usepackage{mathptmx}
\usepackage[numbers,compress]{natbib}
\makeatletter
\journalname{Journal of Low Temperature Physics}
\newcommand{\spider}{{\sc Spider}}
\newcommand{\planck}{{\textsl{Planck}}}

\begin{document}

\newcommand{\hdblarrow}{H\makebox[0.9ex][l]{$\downdownarrows$}-}
\title{Particle response of antenna-coupled TES arrays: results from \spider~and the lab}

\author{B. ~Osherson$^{1}$ \and J.P. ~Filippini$^{1,2}$ \and J. ~Fu$^{2}$ \and R.V. ~Gramillano$^{1}$ \and R. ~Gualtieri$^{1}$ \and E.C. ~Shaw$^{1}$ \and P.A.R. ~Ade$^{3}$ \and M. ~Amiri$^{4}$ \and S.J. ~Benton$^{5}$ \and J.J. ~Bock$^{6,7}$ \and J.R. ~Bond$^{8}$ \and S.A. ~Bryan$^{9}$ \and H.C. ~Chiang$^{10}$ \and C.R. ~Contaldi$^{11}$ \and O. ~Dore$^{6,7}$ \and A.A. ~Fraisse$^{5}$ \and A.E. ~Gambrel$^{12}$ \and N.N. Gandilo$^{13}$ \and J.E. ~Gudmundsson$^{14}$ \and M. ~Halpern$^{4}$ \and J. ~Hartley$^{13}$ \and M. ~Hasselfield$^{15}$ \and G. ~Hilton$^{16}$ \and W. ~Holmes$^{7}$ \and V.V. ~Hristov$^{6}$ \and K.D. ~Irwin$^{17,18}$ \and W.C. ~Jones$^{5}$ \and Z.D. ~Kermish$^{5}$ \and P.V. ~Mason$^{6}$ \and K. ~Megerian$^{7}$ \and L. ~Moncelsi$^{6}$ \and T.A. ~Morford$^{6}$ \and J.M. ~Nagy$^{19,20}$ \and C.B. ~Netterfield$^{13,21}$ \and I.L. ~Padilla$^{21,22}$ \and A.S. ~Rahlin$^{12,23}$ \and C. ~Reintsema$^{16}$ \and J.E. ~Ruhl$^{19}$ \and M.C. ~Runyan$^{7}$ \and J.A. ~Shariff$^{8}$ \and J.D. ~Soler$^{24,25}$ \and A. ~Trangsrud$^{6,7}$ \and C. ~Tucker$^{3}$ \and R.S. ~Tucker$^{6}$ \and A.D. ~Turner$^{7}$ \and A.C. ~Weber$^{7}$ \and D.V. ~Wiebe$^{4}$ \and E.Y. ~Young$^{17,26}$}
\authorrunning{B.~Osherson \textit{et. al}}
\institute{
$^{1}$Department of Physics, University of Illinois at Urbana-Champaign, Urbana, IL, USA \\
$^{2}$('Department of Physics, University of Illinois at Urbana-Champaign, Urbana, IL, USA', 'Department of Astronomy, University of Illinois at Urbana-Champaign, Urbana, IL, USA') \\
$^{3}$School of Physics and Astronomy, Cardiff University, The Parade, Cardiff, CF24 3AA, UK \\
$^{4}$Department of Physics and Astronomy, University of British Columbia, 6224 Agricultural Road, Vancouver, BC V6T 1Z1, Canada \\
$^{5}$Department of Physics, Princeton University, Princeton, NJ, USA 08544, USA \\
$^{6}$Division of Physics, Mathematics and Astronomy, California Institute of Technology, MS 367-17, 1200 E. California Blvd., Pasadena, CA 91125, USA \\
$^{7}$Jet Propulsion Laboratory, Pasadena, CA 91109, USA \\
$^{8}$Canadian Institute for Theoretical Astrophysics, University of Toronto, 60 St. George Street, Toronto, ON M5S 3H8, Canada \\
$^{9}$School of Earth and Space Exploration, Arizona State University, 781 S Terrace Road, Tempe, AZ 85287, USA \\
$^{10}$Department of Physics, McGill University, Ernest Rutherford Physics Building, 3600 rue University, Montreal, QC, Canada H3A 2T \\
$^{11}$Blackett Laboratory, Imperial College London, SW7 2AZ, London, UK \\
$^{12}$Kavli Institute for Cosmological Physics, University of Chicago, 5640 S Ellis Avenue, Chicago, IL 60637 USA \\
$^{13}$Department of Astronomy and Astrophysics, University of Toronto, 50 St George Street, Toronto, ON M5S 3H4 Canada \\
$^{14}$The Oskar Klein Centre for Cosmoparticle Physics, Department of Physics, Stockholm University, AlbaNova, SE-106 91 Stockholm, Sweden \\
$^{15}$Department of Astronomy and Astrophysics, Pennsylvania State University, 520 Davey Lab, University Park, PA 16802, USA \\
$^{16}$National Institute of Standards and Technology, 325 Broadway Mailcode 817.03, Boulder, CO 80305, USA \\
$^{17}$Department of Physics, Stanford University, 382 Via Pueblo Mall, Stanford, CA 94305, USA \\
$^{18}$SLAC National Accelerator Laboratory, 2575 Sand Hill Road, Menlo Park, CA 94025, USA \\
$^{19}$Physics Department, Case Western Reserve University, 10900 Euclid Ave, Rockefeller Building, Cleveland, OH 44106, USA \\
$^{20}$Dunlap Institute for Astronomy, 50 St. George Street, Toronto, Ontario, Canada M5S 3H4 \\
$^{21}$Department of Physics, University of Toronto, 60 St George Street, Toronto, ON M5S 3H4 Canada \\
$^{22}$Department of Physics and Astronomy, Johns Hopkins University, 3400 N. Charles Street, Baltimore, MD 21218 USA \\
$^{23}$Fermi National Accelerator Laboratory, P.O. Box 500, Batavia, IL 60510-5011, USA \\
$^{24}$Max-Planck-Institute for Astronomy, Konigstuhl 17, 69117, Heidelberg, Germany \\
$^{25}$Laboratoire AIM, Paris-Saclay, CEA/IRFU/SAp - CNRS - Universite Paris Diderot, 91191, Gif-sur-Yvette Cedex, France \\
$^{26}$Kavli Institute for Particle Astrophysics and Cosmology, Menlo Park, CA 94025 \\
\email{osherso2@illinois.edu}}

\maketitle

\begin{abstract}

Future mm-wave and sub-mm space missions will employ large arrays of multiplexed Transition Edge Sensor (TES) bolometers. Such instruments must contend with the high flux of cosmic rays beyond our atmosphere that induce `glitches' in bolometer data, which posed a challenge to data analysis from the Planck bolometers. Future instruments will face the additional challenges of shared substrate wafers and multiplexed readout wiring. In this work we explore the susceptibility of modern TES arrays to the cosmic ray environment of space using two data sets: the 2015 long-duration balloon flight of the \spider~ cosmic microwave background polarimeter, and a laboratory exposure of \spider~flight hardware to radioactive sources. We find manageable glitch rates and short glitch durations, leading to minimal effect on \spider~analysis. We constrain energy propagation within the substrate through a study of multi-detector coincidences, and give a preliminary look at pulse shapes in laboratory data.

\keywords{cosmic microwave background, transition edge sensor, bolometer, cosmic ray}

\end{abstract}

\section{Introduction}
Modern instruments characterizing the cosmic microwave background (CMB) typically employ large-scale integrated arrays of bolometric sensors to achieve high sensor counts at high per-sensor sensitivities~\cite{S4Tech2017}. A typical array consists of hundreds or thousands of membrane-isolated superconducting transition-edge sensors (TESs)~\cite{IrwinHilton2005} photolithographically patterned onto a common silicon wafer, each coupled to incident radiation via feedhorns~\cite{Britton:2010qx}, lenses~\cite{Arnold:2012sr}, or synthesized antennas~\cite{JPLdet2015}. Integrated bolometer arrays are a key enabling technology for a number of proposed space-based astronomical instruments~\cite{2018LiteBIRD,SPICA2018,OSTinterim2018,PICOreport2019}.

Though designed to detect electromagnetic radiation, such bolometers respond to any change in the thermal balance of their temperature sensors. In particular, cosmic rays --- energetic particles from space --- induce sharp `glitches' in bolometer data streams that complicate astrophysical observations with balloon- and satellite-borne instruments. A recent example comes from the \planck~satellite's bolometer-based High-Frequency Instrument (HFI)~\cite{PlanckMission2011,PlanckOverviewLegacy2018}. The high rate ($\sim$1~Hz) and long recovery times (as long as $\sim$2~s) of cosmic ray glitches led to significant analysis challenges and data loss~\cite{Planck2013_X_CosmicRay}. Careful analyses of flight and laboratory data~\cite{Catalano:aap2014,Catalano:jltp2014,Planck2013_X_CosmicRay} led to successful cosmological analysis and elucidated several glitch classes corresponding to energy depositions in different regions of the detector assembly. Cosmic ray response is thus a key performance consideration for future space-based bolometric instruments, which will face the added complications of large shared wafers and crosstalk within multiplexed readout wiring.

In this work we explore the susceptibility of modern TES arrays to the cosmic ray environment of space using two data sets: the 2015 long-duration balloon flight of \spider~\cite{Filippini2010,Fraisse2013,Nagy2017} and laboratory exposure of \spider~flight hardware to radioactive sources. Our goal is a system-level characterization of a full detector/readout assembly to inform expectations and designs for future missions.

\section{Data sets}
\begin{figure}[tbp]
\centering
\begin{minipage}[ht]{.4\linewidth}
\begin{center}
    \includegraphics[width=\textwidth]{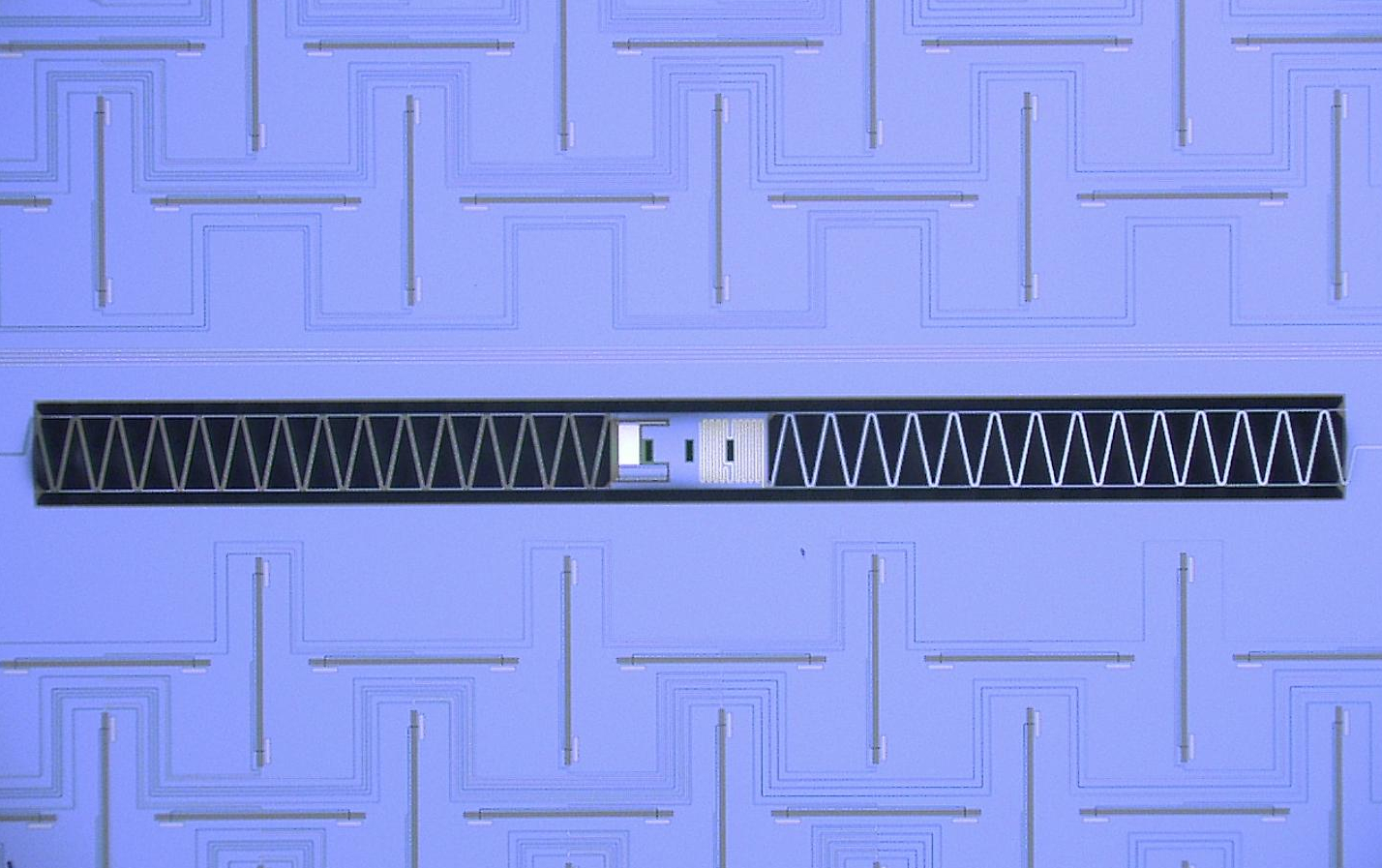}
\end{center}
\end{minipage}\hfill
\begin{minipage}[ht]{.5\linewidth}
\begin{center}
    \includegraphics[width=\textwidth]{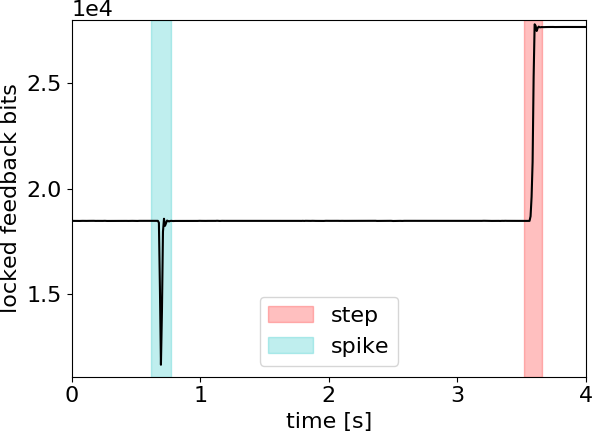}
\end{center}
\end{minipage}
\caption{{\bf A} \textit{(Left)}: Optical image of a single \spider~bolometer, showing the suspended island, the meandered isolation legs, and the slot antenna network etched into the surrounding Nb ground plane. {\bf B} \textit{(Right)}: An interval of lab data showing the two major glitch classes. Time stream noise is of order $\sim$1~bit r.m.s.}
\label{example_and_zoom}
\vspace{-15pt}
\end{figure}

\spider~\cite{Filippini2010,Fraisse2013,Nagy2017} is a balloon-borne instrument designed to measure the $B$-mode signature of primordial gravitational waves in the CMB. \spider's first long-duration balloon flight in January 2015 observed the sky for 16 days from an altitude of 37~km over Antarctica. The payload employed six refracting telescopes (three each at 94 and 150~GHz), each illuminating a 300~mK focal plane populated with JPL antenna-coupled TES arrays~\cite{JPLdet2015} (Fig.~\ref{example_and_zoom} A). Data from \spider's 2400 TESs are recorded to disk at 119~Hz using a three-stage time-division SQUID multiplexer system~\cite{deKorte2003,Stiehl2011} managed by UBC's Multi-Channel Electronics (MCE)~\cite{Battistelli2008}. 
\spider's flight thus provides a useful system-level proxy for a modern bolometer array and readout system in a space-like cosmic ray environment. We discuss observations from these data in Section~\ref{SPIDER_res}.

After \spider's successful flight, we constructed a dedicated test stand at the University of Illinois to expose a single detector wafer to an arrangement of radioactive sources. The data here were taken with a single detector wafer (128 TESs) and associated multiplexer hardware and sub-Kelvin cooler, all recovered from \spider's flight payload. The wafer was exposed to radiation from four localized Am-241 sources, chosen to recreate high energy depositions while allowing more control over event location and data acquisition. These data are discussed in Section~\ref{Lab_res}.

For both flight and laboratory data, analysis consists of three major steps: flagging glitches, estimating their energies, and identifying `coincidences' --- events in which multiple detectors experience glitches simultaneously. In both data sets we identify two types of glitches (Fig.~\ref{example_and_zoom} B): \emph{spikes}, which recover to the pre-glitch baseline within only a few samples at the flight data rate, and \emph{steps}, which introduce a persistent offset in the data by one flux quantum of the first-stage SQUID. For spike glitches we estimate the apparent deposited energy on the TES island from the integrated change in TES Joule power during the glitch --- a good proxy for deposited energy in the limit of small signal and strong electrothermal feedback~\cite{IrwinHilton2005}. While we cannot estimate the individual event energies for step glitches, the model described in Section \ref{Glitch_shapes} suggests that they are relatively high ($\gtrsim$100~keV).

To calibrate expectations, we make a crude estimate of the cosmic ray response of a \spider\ detector array by combining \planck's estimated cosmic ray flux ($\sim$5/cm$^2$/s above 40~MeV~\cite{Planck2013_X_CosmicRay}) and the stopping power of a minimum-ionizing particle. This suggests typical energy depositions in the bolometer island of order a few hundred eV every $\sim$10 minutes, with perhaps a similar rate on the suspension legs; a more detailed Monte Carlo model incorporating realistic incident energies and particle showers is left to future work. Interactions in the larger, thicker wafer should be far more frequent ($\sim$250~Hz) and energetic (many keV to MeV). A central question is thus the distance over which each TES responds significantly to wafer interactions, via wafer temperature excursions or out-of-equilibrium ballistic phonon propagation. If this is large, we would expect a high event rate and frequent coincident events among nearby bolometers.

\section{Flight analysis}
\label{SPIDER_res}
At the low sampling rate of \spider\ flight data, the shapes of both steps and spikes are essentially determined by the MCE's digital anti-aliasing filter. Spikes are identified using a simple search for isolated samples that deviate from the data mean by $>5\sigma$, while steps are identified using a matched filter. Spike energies are estimated in two ways: by direct pulse integration, and via the use of a matched filter (the `template estimator') that is calibrated to match the pulse integral at $\sim$keV energies. The latter gives superior energy resolution, at the cost of less accuracy for energetic events in which TES saturation distorts pulse shape. In flight data the present analysis imposes thresholds between 200-500~eV, which might be reduced with greater care to data cleaning.

After excluding periods of known poor detector performance identified for the B-mode analysis, this pipeline identifies spike glitches about one every three minutes in each detector, and step glitches roughly once per hour per detector. Step glitches are generated more frequently during known periods of intermittent radio-frequency interference (excluded from analysis). This rate of glitches is broadly comparable to (though somewhat higher than) our na\"ive expectations for the island hit rate, suggesting that each bolometer is sensitive to interactions over an area within a factor of a few of that of the suspended island. Given the observed low rates and brief durations, cosmic ray glitches have negligible impact on \spider's science analysis; A conservative interval of data is excised around each identified glitch or step and the resulting gap is filled with simulated noise. For steps, the two baselines are also aligned across the gap.

\begin{figure}[tbp]
\centering
\begin{minipage}[ht]{.475\linewidth}
\begin{center}
    \includegraphics[width=\textwidth]{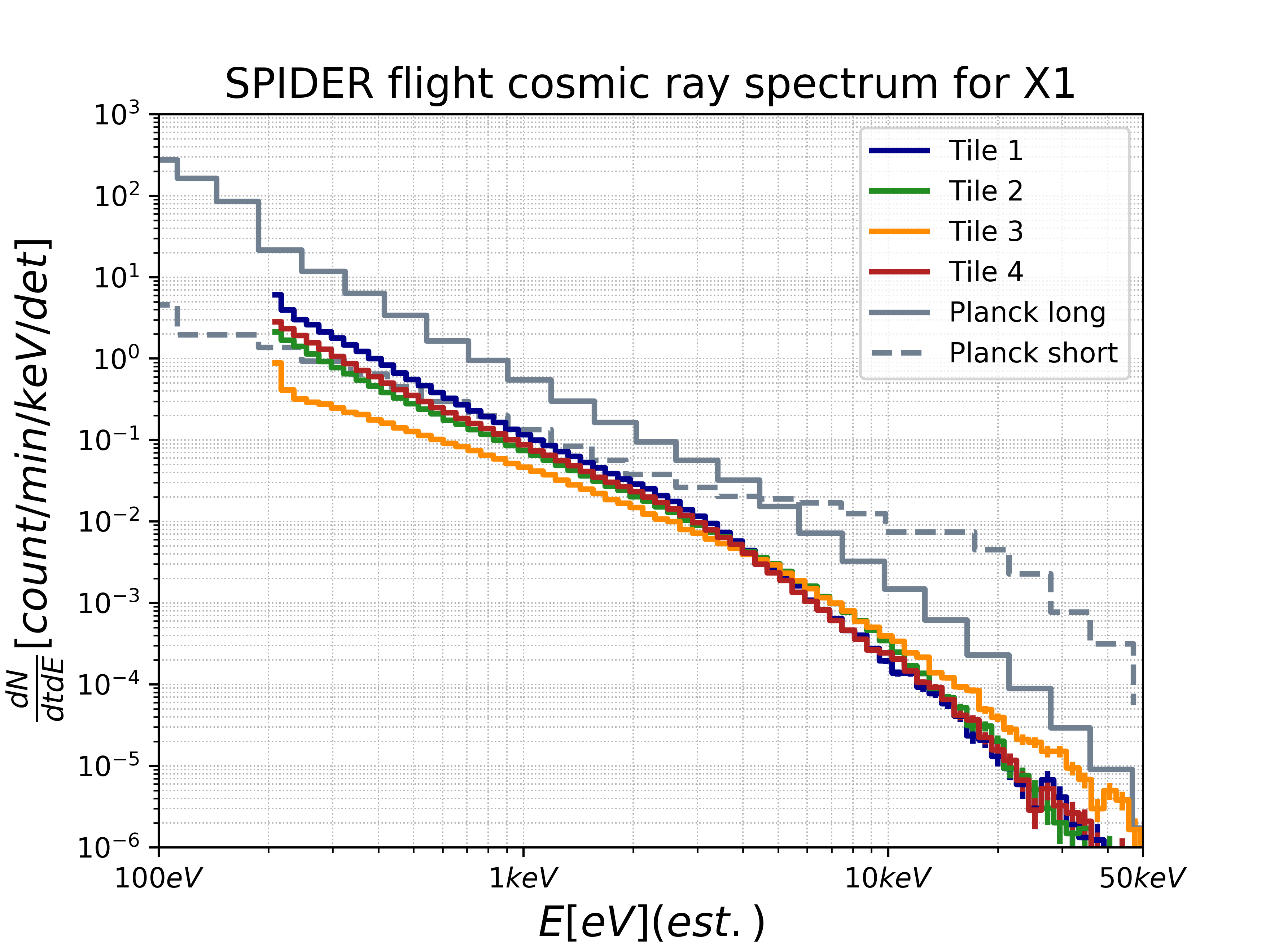}
\end{center}
\end{minipage}\hfill
\begin{minipage}[ht]{.525\linewidth}
\begin{center}
    \includegraphics[width=\textwidth]{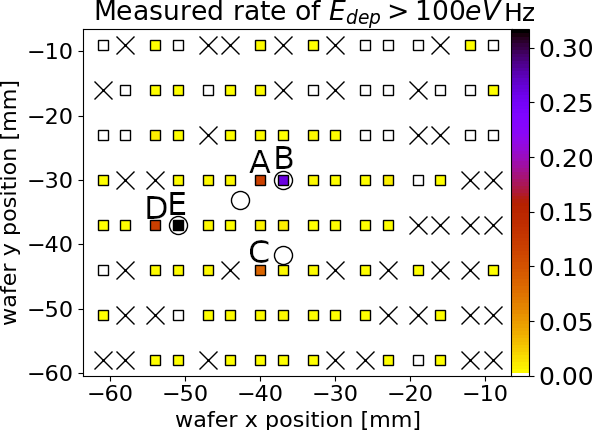}
\end{center}
\end{minipage}
    \caption{{\bf A} \textit{(Left)}: Flight data spectra for each of four wafers in one 150~GHz telescope. Included are similar Planck spectra from ~\cite{Catalano:jltp2014}. {\bf B} \textit{(Right)}: Arrangement of bolometers (\emph{squares}) and sources (\emph{circles}) in our Am-241 test. Detectors labeled B and E are directly under sources. Detectors A and D, which also show elevated rates, are the physical and multiplexing neighbors of B and E, respectively. Detector C, also elevated, is the other multiplexing neighbor of B, and so subject to known crosstalk. Other detectors (even those near sources) show low event rates, suggesting that detectable energy does not propagate beyond a few millimeters. X's indicate detectors excluded by a quality cut. The rates do not differ significantly between the two energy estimators at these energies.}
    \label{spectra}
\end{figure}

A representative spectrum of spike energies from one of the six telescopes is shown in Fig.~\ref{spectra} A. Most detectable glitches are a few hundred eV or less, piling up at the edge of each detector's analysis threshold. Note that not all detector wafers are in perfect agreement of the spectrum of energy depositions; we are exploring possible sources of this variability in detector properties and analysis for discussion in future work. 

After compiling a catalog of glitches for each detector, we identify glitches occurring on different detectors within $\pm$1 sample of one another. 
Across all SPIDER wafers $\sim$5\% of glitches are in coincidence with a glitch in a different detector, and $\sim$90\% of these coincidences feature only two detectors. Rare high-multiplicity events are also seen, some of which may be caused by thermal fluctuations or showering from a cosmic ray collision in \spider's hull.
We observe clear excesses of coincident events above random chance among two classes of detector pairs. The first is detectors within a `polarization pair': channels whose slot antennas are interleaved on the same patch of wafer. Polarization pairs are both physically closer than other bolometers ($\sim$1~mm) and `multiplexing neighbors', read out consecutively by the same amplifier during the multiplexing sequence. The second pairing of interest are multiplexing neighbors which are not physical neighbors, typically a few centimeters away. The polarization pairs have a measured rate of coincident glitches $\sim20\times$ higher than for detector pairings that are neither physical nor multiplexing neighbors, while the distant multiplexing neighbor has a measured coincidence rate $\sim3-4\times$ higher than non-neighbor pairings. The excess rate of coincidences may be the result of the known $\sim0.3$\% inductive cross-talk between neighboring channels in the multiplexing scheme~\cite{deKorte2003} (glitch crosstalk will only be detectable from rarer high-energy depositions), by a narrow shower of particles ejected by a cosmic ray collision elsewhere, or by energy propagating from deposition in the wafer to the nearest bolometers. The lack of coincidences among more distant neighbors suggests that detectable energy propagation, if present, is constrained to length scales of millimeters rather than centimeters.

\section{Laboratory analysis}
\label{Lab_res}

\begin{figure}[tbp]
\centering
\begin{minipage}[ht]{.425\linewidth}
\begin{center}
    \includegraphics[width=\textwidth]{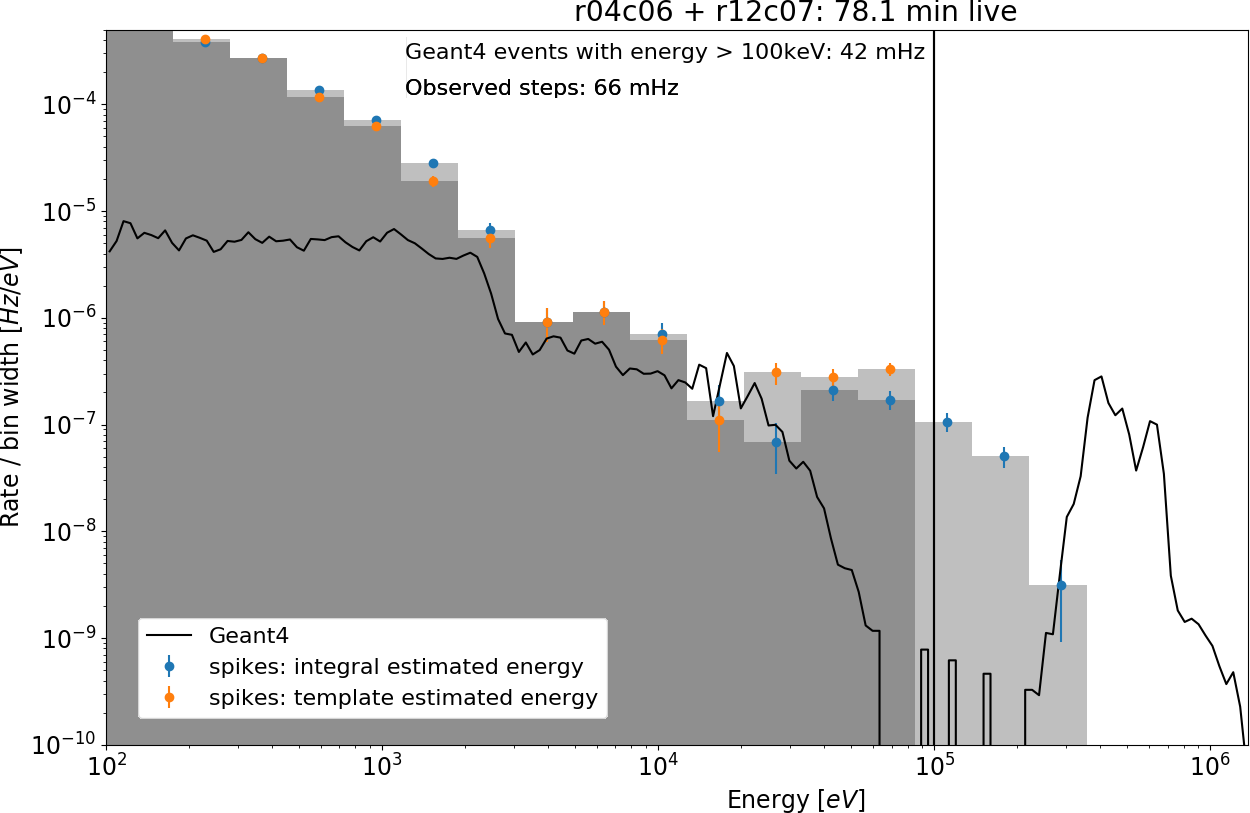}
\end{center}
\end{minipage}\hfill
\begin{minipage}[ht]{.575\linewidth}
\begin{center}
    \includegraphics[width=\textwidth]{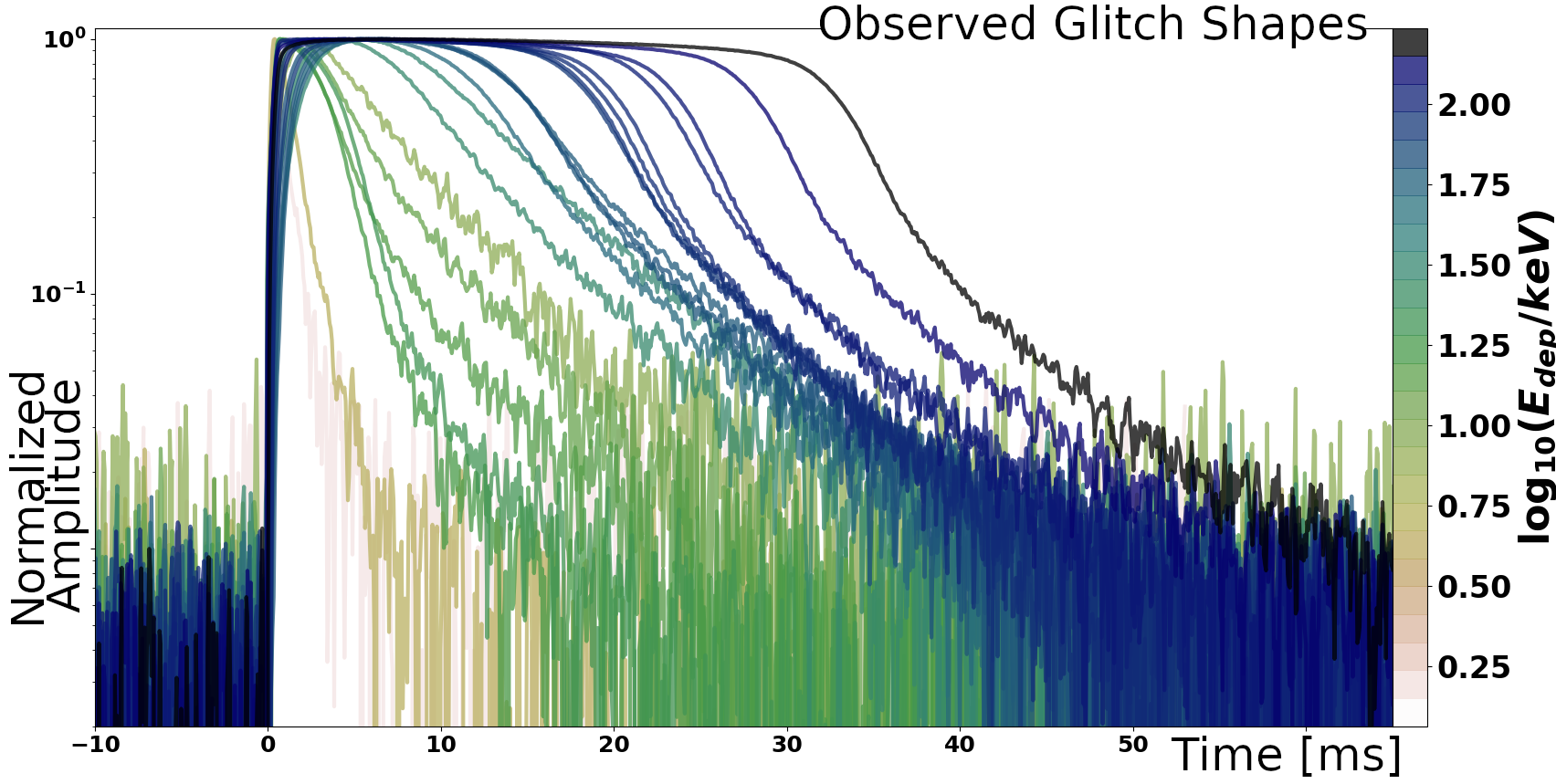}
\end{center}
\end{minipage}
\caption{{\bf A} \textit{(Left)}: The combined glitch catalog of two detectors, labeled `B' and `E' in Fig.~\ref{spectra} B, both directly beneath the Am-241 sources, is converted into two spectra: one using the integral energy estimator and the other using the template energy estimators. The Geant4 simulated spectrum of energy depositions directly into a TES island (solid black line) is added for comparison. Annotations show the measured rate of step glitches (not included in the historgrams).
{\bf B} \textit{(Right)}: A variety of spike glitches sampled at 15 kHz, amplitude-normalized and aligned  with energies from the integral estimator. TES saturation is visible as flat peaks.}
\label{lab_data}
\vspace{-15pt}
\end{figure}

In order to better explore the localization and pulse shapes of particle events, we exposed a \spider~wafer to radioactive Am-241, a 5.5~MeV alpha emitter. We place four 0.9~$\mu$Ci sources behind pinhole collimators positioned over the bolometer/wiring (non-sky) face of the detector wafer. The source arrangement is shown in Fig.~\ref{spectra} A: two of the sources are placed directly over bolometer islands, while the other two sources are placed over the wafer far from any TES. The collimated spot size is somewhat larger than the island cutout, so we expect a high rate of alpha hits to the surrounding wafer even for the island sources. The alpha particle is expected to deposit $\sim$400~keV in the island and be fully stopped by the $\sim$400~$\mu$m silicon wafer, both well in excess of typical flight cosmic ray energy depositions in these components. 
Glancing interactions, gamma rays, and ejected electrons will further contribute a wide spread of deposited energies. The analysis is very similar to that in Section~\ref{SPIDER_res}, but with somewhat lower thresholds (as low as 50~eV, as measured by the template estimator). Long data sets were acquired on all detectors, with similar sampling rates and filtering as for the flight data, as well as shorter data sets sampling subsets of detectors at higher rates (15~kHz and 250~kHz, without digital filtering) for closer examination of pulse shapes (Fig.~\ref{lab_data} B). Out of transition or pathological detectors are removed from the analysis and are shown with `X's in Fig.~\ref{spectra} A. Some data in the remaining detectors are masked in the case of ringing from our regular resetting of the digital filter's buffers. As in flight, the live time of each detector --- used to compute the measured rate of interactions --- is updated to reflect all data cuts. 

For comparison with data incorporating these effects we carried out Monte Carlo simulations of the source, TES island, and surrounding materials using the Geant4 package~\cite{Geant4:2003,Geant4:2006,Geant4:2016}. 
The simulated TES island model incorporates the various patterned metal and dielectric layers constructed as a simplified set of multi-layer `stacks' of appropriate areas and thicknesses. 
In addition to the expected peak from alpha particle interactions, a power-law-like spectrum of energies is visible from gamma rays and secondary electrons. 
We model events farther from the collimator using ensembles of simulations in which each of these stacks is expanded to the area of the whole wafer and statistics are accrued as a function of radius from the collimator.
Even for TES islands relatively near the source, the simulated rate is two orders of magnitude lower than for islands directly beneath the source. Not simulated is the irreducible glitch rate induced by cosmic ray interactions, which should yield only a handful of events in these data.

Fig.~\ref{lab_data} A compares the observed and simulated energy spectra for a representative TES directly beneath a collimated source. There are clear discrepancies: the alpha particle peak ($\gtrsim100$~keV) is absent or suppressed in energy, while the data show an excess of low-energy events ($\lesssim1$~keV). We postulate that the former may appear instead as step discontinuities (see Section \ref{Glitch_shapes}), while the latter may be caused by simulation deficiencies or by alpha energy depositions in the suspension legs or nearby wafer, from which only a fraction of the deposited energy reaches the bolometer (to be modeled in future work). 

Fig.~\ref{spectra} B shows the glitch rate in a 40-minute laboratory source data set. Glitches are identified above detector-dependent energy thresholds, excluding non-functional and noisy channels. As expected, the two detectors directly under the sources (`B' and `E') show the highest glitch rates by far. These rates are $\sim5\times$ larger than predicted by the Geant4 simulations. Though some of this may be ascribed to uncertainties in source flux through the collimator and simulation completeness at the lowest energies, this is consistent with measurable response to interactions in an area perhaps modestly larger than the physical island. Nearly all other detectors see a very low rate of glitches, including those near to sources illuminating antenna patches.

Elevated event rates are seen for a small number of detectors that are adjacent to those under the source, either physically and/or in multiplexing sequence (see figure caption). For detector `C', which is physically distant from a source, the excess is likely attributable to known multiplexer crosstalk: the overwhelming majority of its events are coincident with events in its multiplexing neighbor `B', which is beneath a source. The two polarization partners of the source-illuminated detectors, which are both physical and multiplexing neighbors, show yet higher event rates suggestive of additional energy propagation beyond crosstalk. Events in one of these (`D') are overwhelmingly coincident with those in its partner. For the other (`A'), however, $\sim$65\% of events are not coincident with its partner. This difference is not currently understood, but may be due to source alignment. Overall, lab data is qualitatively consistent with flight analysis: TES channels respond to interactions within at most a few millimeters, and via crosstalk to large glitches on their multiplexing neighbors.

\section{Glitch shapes}
\label{Glitch_shapes}
In addition to the energy and coincidence analyses above, laboratory data taken at elevated sample rates ($\sim$15~kHz and 250~kHz) have informed our understanding of the glitch shapes:
\begin{itemize}
\item \emph{Step glitches}: 
The MCE uses an integral feedback loop to linearize the response of the SQUID readout chain. Sufficiently large glitches (amplitudes $\gtrsim\Phi_0/2$ at the first-stage SQUID) can cause this feedback controller to lose lock on the glitch's fast rising edge, slipping to a lock point one flux quantum away as the glitch recovers. This phenomenon is visible in high-rate laboratory data and has been simulated in a computer model that includes TES, SQUIDs, and feedback control algorithm. Data and simulations show that the glitch/step energy threshold varies with bias current, and that step events are absent at high enough biases where TES saturation occurs before a flux slip is triggered.

\item \emph{Spike glitches}: 
High-rate studies of ordinary spike glitches show a range of time constants for the rising and falling pulse edges as well as TES saturation at high energies (appearing as plateaus in half of the glitches) represented in Fig.~\ref{lab_data} B. The remaining multiplicity of time constants in some of the individual glitches, as well as the inconsistency of these time constants from glitch to glitch is a subject for future laboratory tests and modeling, but may be related to energy depositions in different portions of the bolometer and suspension structure.
\end{itemize}

\section{Conclusions}
We have exposed a modern antenna-coupled TES array and readout system to particle radiation from cosmic rays and radioactive sources. Event rates and coincidence analyses suggest a limited impact from particle interactions in the shared wafer substrate, likely confined to a few millimeters surrounding the suspended TES itself. We have identified the origin of the `flux slip' phenomenon --- a multiplexer effect in which large energy depositions induce step discontinuities in our TES data. Overall, cosmic ray glitches are expected to have minimal impact on the science analysis of \spider~data. 
In the future we expect to expand this work to devices with lower operating temperatures ($T_c\sim 100$~mK), where phonon physics and noise levels will more closely resemble those of future space instruments, alongside more detailed simulations of the bolometer thermal architecture and particle environment.

\begin{acknowledgements}
This work is supported by NASA’s Strategic Astrophysics Technology program (14-SAT14-0009, 16-SAT16-0002). SPIDER is supported by in the U.S.A. by NASA (NNX07AL64G, NNX12AE95G, NNX17AC55G) and NSF (PLR-1043515); in Canada by NSERC and CSA; as well as by the Research Council of Norway, the Swedish Research Council, and the Packard Foundation. Logistical support in Antarctica is supported by the NSF through the U.S. Antarctic Program. The collaboration is grateful to the British Antarctic Survey, particularly Sam Burrell, for invaluable assistance with data and payload recovery after the 2015 flight.
\end{acknowledgements}


\begin{thebibliography}{99}

\bibitem{S4Tech2017}
Abitbol, M. H., Ahmed, Z., Barron, D., et al., arXiv:1706.02464 (2017).

\bibitem{IrwinHilton2005}
Irwin, K.~D., \& Hilton, G.~C., "Transition-Edge Sensors" in Cryogenic Particle Detection, Topics in Applied Physics, vol. 99, Ed. C. Enss, Springer (2005)

\bibitem{Britton:2010qx}
Britton, J. W., Nibarger, J. P., Yoon, K. W., et al., Proc. SPIE Int. Soc. Opt. Eng., 7741, 21 (2010), doi: 10.1117/12.857885

\bibitem{Arnold:2012sr}
Arnold, K., et al., Proc. SPIE Int. Soc. Opt.
Eng., 8452, 84521D (2012), doi: 10.1117/12.927057

\bibitem{JPLdet2015}
BICEP2 Collaboration; Keck Array Collaboration; SPIDER Collaboration, Ade, P.A.R., Aikin, R.W., Amiri, M. et al, {\it The Astrophysical Journal}, \textbf{812}, 17, (2015), DOI: 10.1088/0004-637X/812/2/176  

\bibitem{2018LiteBIRD}
Hazumi, M., Ade, P. A. R., Akiba, Y., et. al., {\it J. Low Temp. Phys.} \textbf{194}, 5, 443 (2019).

\bibitem{SPICA2018} Roelfsema, P., Shibai, H., Armus, L., et al., {\it Publications of the Astronomical Society of Australia}, {\bf 35}, E030 (2018). doi:10.1017/pasa.2018.15

\bibitem[The OST mission concept study team(2018)]{OSTinterim2018} The OST mission concept study team, arXiv:1809.09702 (2018).

\bibitem[Hanany et al.(2019)]{PICOreport2019} Hanany, S., Alvarez, M., Artis, E., et al., arXiv:1902.10541 (2019).

\bibitem[Planck Collaboration et al.(2011)]{PlanckMission2011} Planck Collaboration, Ade, P.~A.~R., Aghanim, N., et al., {\it Astronomy \& Astrophysics}, \textbf{536}, A1 (2011).

\bibitem[Planck Collaboration et al.(2018)]{PlanckOverviewLegacy2018} Planck Collaboration, Akrami, Y., Arroja, F., et al.\, arXiv:1807.06205 (2018). 

\bibitem{Planck2013_X_CosmicRay}
Planck Collaboration, Ade, P.A.R., Aghanim, N., Arnaud, M. et al., {\it Astronomy \& Astrophysics}, , \textbf{571} , A10, (2014), DOI: 10.1051/0004-6361/201321577 

\bibitem{Catalano:aap2014}
Catalano, A., et al., \textit{Astron. Astrophys.} \textbf{569}, A88 (2014), doi: 10.1051/0004-6361/201423868

\bibitem{Catalano:jltp2014}
Catalano, A., Ade, P., Atik, Y. et al., {\it Journal of Low Temperature Physics}, \textbf{176}, 5, pp. 773-786, (2014), DOI: 10.1007/s10909-014-1116-6

\bibitem{Filippini2010}
Filippini, J.~P., Ade, P.~A.~R., Amiri, M., et al., {\it Proc. SPIE}, \textbf{7741}, (2010), DOI: 10.1117/12.857720

\bibitem{Fraisse2013}
Fraisse, A.~A., Ade, P.~A.~R., Amiri, M., et al., {\it Journal of Cosmology and Astroparticle Physics}, \textbf{4}, 047 (2013)

\bibitem{Nagy2017}
Nagy, J.~M., Ade, P.~A.~R., Amiri, M., et al., {\it Astrophys. J}, \textbf{844}, 151 (2017), DOI: 10.3847/1538-4357/aa7cfd 

\bibitem{deKorte2003}
de Korte, P.~A.~J., Beyer, J., Deiker, S., et al., Review of Scientific Instruments, 74, 3807 (2003).

\bibitem{Stiehl2011}
Stiehl, G.~M., Cho, H.~M., Hilton, G.~C., et al., IEEE Transactions on Applied Superconductivity, 21, 298 (2011).
\bibitem{Battistelli2008}
Battistelli, E.~S., Amiri, M., Burger, B., et al., Journal of Low Temperature Physics, 151, 908 (2008).

\bibitem{Geant4:2003}
Agostinelli, S., Allison, J., Amako, K., Apostolakis, J., et al., Nucl. Instr. Meth. A \textbf{506}, 3, 250 (2003).
\bibitem{Geant4:2006}
Allison, J., Amako, K., Apostolakis, J., Araujo, H., et al., IEEE Trans. Nucl. Sci. \textbf{53}, 1, 270 (2006).
\bibitem{Geant4:2016}
Allison, J., Amako, K., Apostolakis, J., Arce, P., et al., Nucl. Inst. Meth. A \textbf{835}, 186 (2016).

\bibitem{Perinati2010}
Perinati, E., Mineo, T., Colasanti, L. et al., {\it Proc. SPIE}, \textbf{7732}, (2010), DOI: 10.1117/12.856393 




\end{thebibliography}
\end{document}